\documentclass[twocolumn,preprintnumbers,amsmath,amssymb,showkeys]{revtex4}

\usepackage{graphicx}
\usepackage{dcolumn}
\usepackage{bm}

\begin{document}
\title{Scale-rich metabolic networks: background and introduction}

\author{Reiko Tanaka}
 \email{reiko@bmc.riken.jp}
\affiliation{%
Bio-Mimetic Control Research Center, RIKEN,
Moriyama-ku, Nagoya 463-0003, Japan
}%
\author{John Doyle}
\affiliation{
Control and Dynamical Systems
California Institute of Technology,
 MC 107-81, Pasadena, CA 91125, USA
}%

\date{\today}

\begin{abstract}
Recent progress has clarified many features of the global
architecture of biological metabolic networks, which have
highly organized and optimized tolerances and tradeoffs
(HOT) for functional requirements of flexibility,
efficiency, robustness, and evolvability, with constraints
on conservation of energy, redox, and many small moieties.
One consequence of this architecture is a highly structured
modularity that is self-dissimilar and scale-rich, with
extremes in low and high
variability, including power laws, in both metabolite and
reaction degree distributions.  This paper illustrates
these features using the well-understood stoichiometry of
metabolic networks in bacteria, and a simple model of an
abstract metabolism.
\end{abstract}

\keywords{Metabolic networks, s-graph, degree distribution,
function modules}
\maketitle

\begin{figure*}[t]
  \begin{center}
    \includegraphics[width=0.8\linewidth]{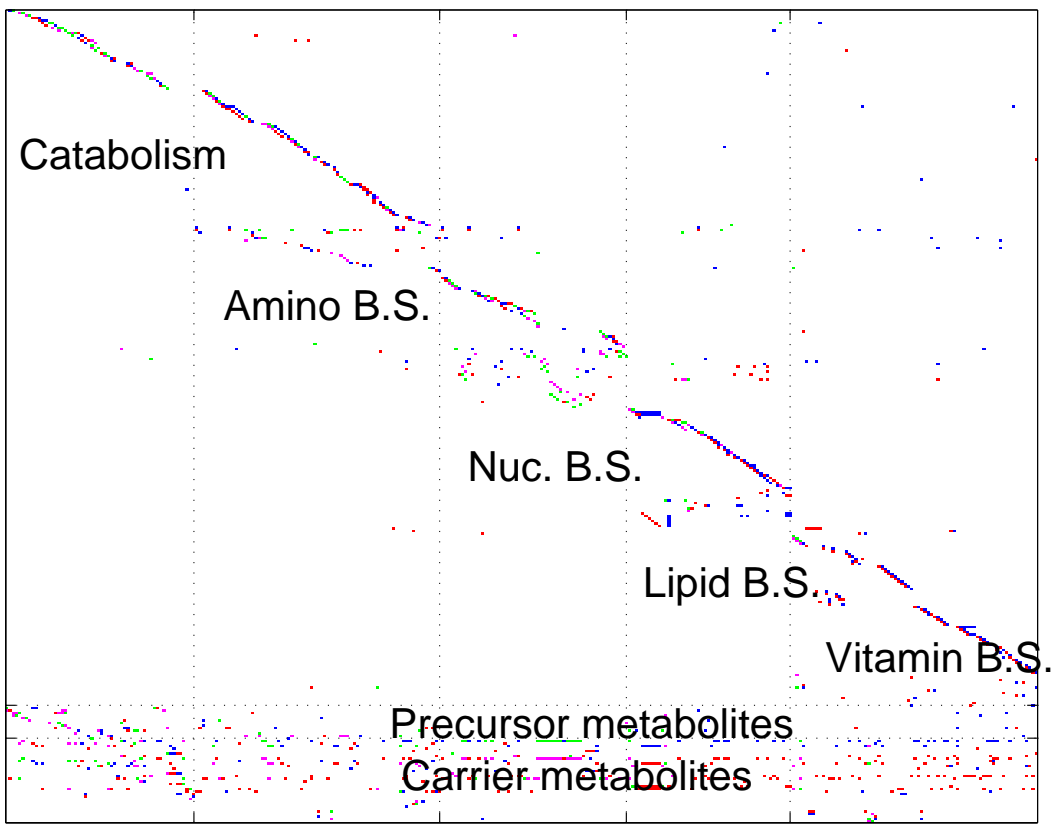}
  \end{center}
 \caption{
S-matrix for {\it H.Pylori} metabolism of with 325
metabolites and 315 reactions, with functional
decomposition into modules. Red and blue correspond to
positive and negative elements, respectively, for
irreversible reactions, and pink and green correspond to
positive and negative elements, respectively, for
reversible reactions. Reactions (columns) have standard
functional modules of catabolism and biosynthesis, which is
further split into amino acid, nucleotide, fatty
acid/lipid/cell structures, and cofactor biosynthesis. The
rows (metabolites) are arranged by their role in reaction
modules to clarify the sparsity pattern of long chains of
successive reactions from inputs to outputs in each module.
The bottom rows are precursor metabolites and carrier
metabolites, which appear throughout different reaction
modules. }
  \label{fig:whole_mat}
\end{figure*}

The simplest model of metabolic networks is a stoichiometry
matrix, or s-matrix for short, with rows of metabolites and
columns of reactions. For example, for the two reactions
{\small
\begin{equation}\label{ichi}
\left. {\begin{array}{*{20}c}
   {S_1  + ATP \to S_2  + ADP},\\
   {S_3  + NADH \to S_4  + NAD}, \\
\end{array}\quad } \right\}
\end{equation}
} \noindent among four carriers ATP, ADP, NADH and NAD and
four other metabolites $S_1, S_2, S_3, S_4$, the s-matrix
is given by {\small
\begin{equation}
\left[
\begin{array}{rr rr rr rr}
-1 & 1 & 0 & 0 & -1 & 1 & 0 & 0\\
0 & 0 & -1 & 1 & 0 & 0 & -1 & 1
\end{array}
 \right] ^{\rm T}.
\label{stoich_mat} %
\end{equation}
} \noindent Metabolic stoichiometry is perhaps the most
unambiguously known aspect of biological networks and makes
an attractive basis for contrasting different approaches to
complex networks \cite{Bilke,NetBio,book,Jeong,Hie}. Figure
\ref{fig:whole_mat} shows a color-coding of the s-matrix
for {\it H.Pylori} core metabolism \cite{data} (all
conclusions are essentially the same for the larger
s-matrix of {\it E. Coli}), with both metabolites and
reactions decomposed into modules. The function of each
reaction module is to make output metabolites from input
metabolites.  For example, catabolism takes external
nutrients and activates carriers and makes 13 precursor
metabolites, and amino acid biosynthesis outputs amino
acids with these precursor metabolites as inputs.
Metabolites are categorized into precursor, carrier, and
other (than precursor and carrier) metabolites. Precursor
metabolites are outputs of catabolism and are the starting
points for biosynthesis. Carrier metabolites correspond to
conserved quantities and are activated in catabolism and
act as carriers to transfer energy and phosphate groups, 
hydrogen/redox, amino groups, acetyl groups, one carbon units
throughout all modules. 
The list of the carrier
metabolites considered here is shown in Table \ref{carriers} in appendix.
 (Some metabolites act
as carriers only in some reactions, but not in others.)
The other (than carriers and precursors) metabolites occur
primarily in separate reaction modules.

\begin{figure}[h]
  \begin{center}
    \includegraphics[width=0.5\linewidth]{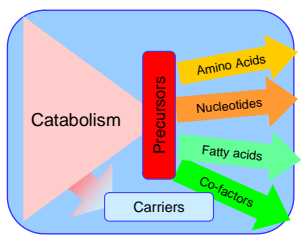}
\end{center}
  \caption{
Schematic drawing of global "bow-tie" structure in general
metabolic networks. }
  \label{fig:bow-tie}
\end{figure}

This categorization of metabolites is compatible with the
standard schematic  `bow-tie' structure of metabolism as
shown in Fig.~\ref{fig:bow-tie}, where a large `fan-in' of
nutrient inputs is catabolized to produce a small handful
of activated carriers and precursor metabolites, which then
`fan-out' to the biosynthesis of a large number of primary
building blocks \cite{Marie2}. The biologically natural
modular decomposition in metabolites is thus into `knot`
(carriers and precursors) and non-`knot' (others) parts of
the `bow-tie.' Further examination of this structure of
stoichiometry shows it to facilitate a variety of highly
organized and optimized tolerances and tradeoffs (HOT)
\cite{CD2} for flexibility, adaptability, efficiency,
robustness, and evolvability\cite{Marie1, Marie2} in the
face of a large number of constraints on conserved
quantities. Thus it is an architecture that is ubiquitous
throughout biology and advanced technologies as well. While
this is all a network-level interpretation of standard
textbook biochemistry, statistical studies \cite{Ma2} of
$80$ fully sequenced organisms produces similar conclusions
about the universal `bow-tie' structure of metabolism.

\begin{figure}[h]
  \begin{center}
    \includegraphics[width=0.23\linewidth]{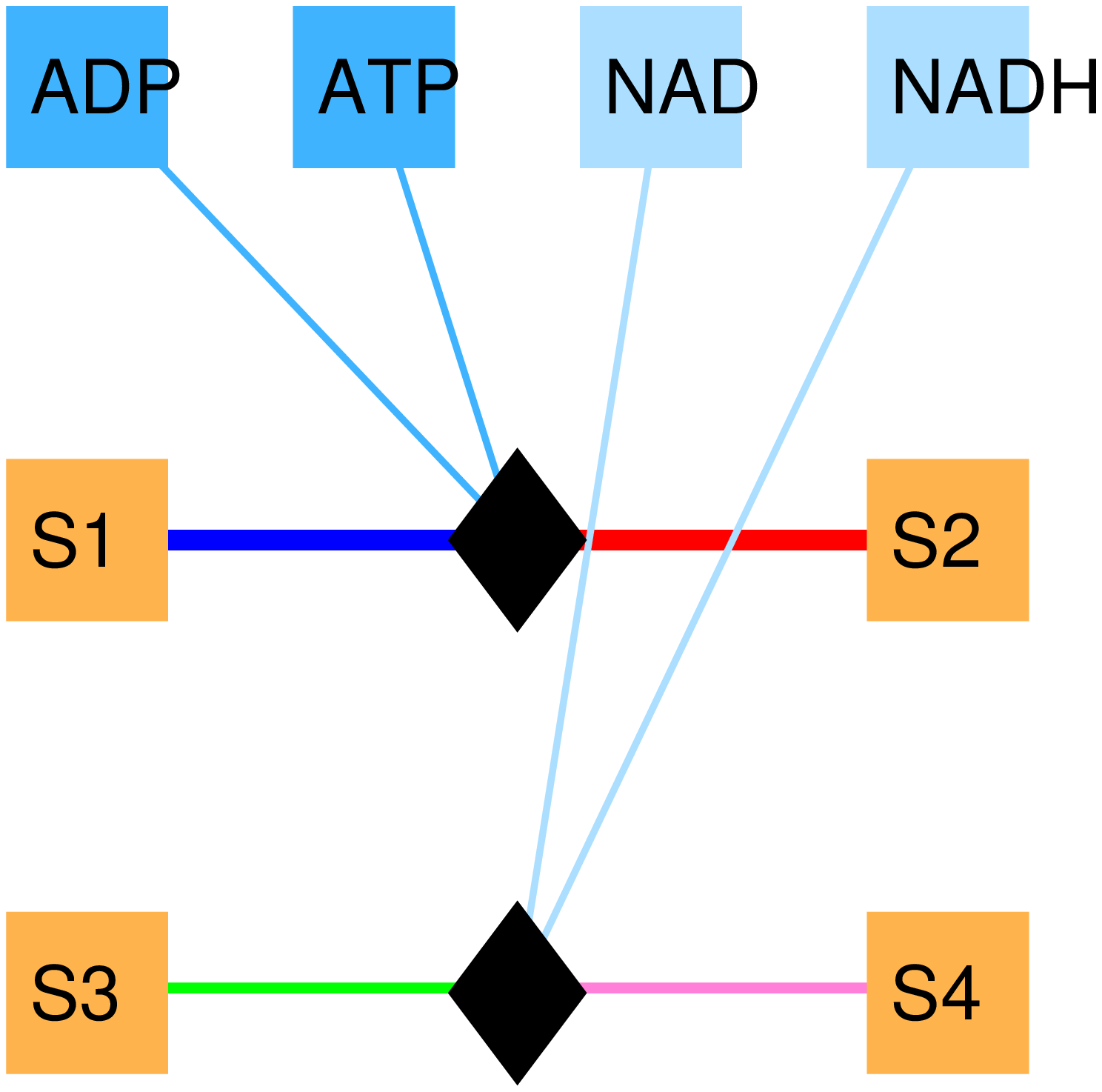}
    \includegraphics[width=0.25\linewidth]{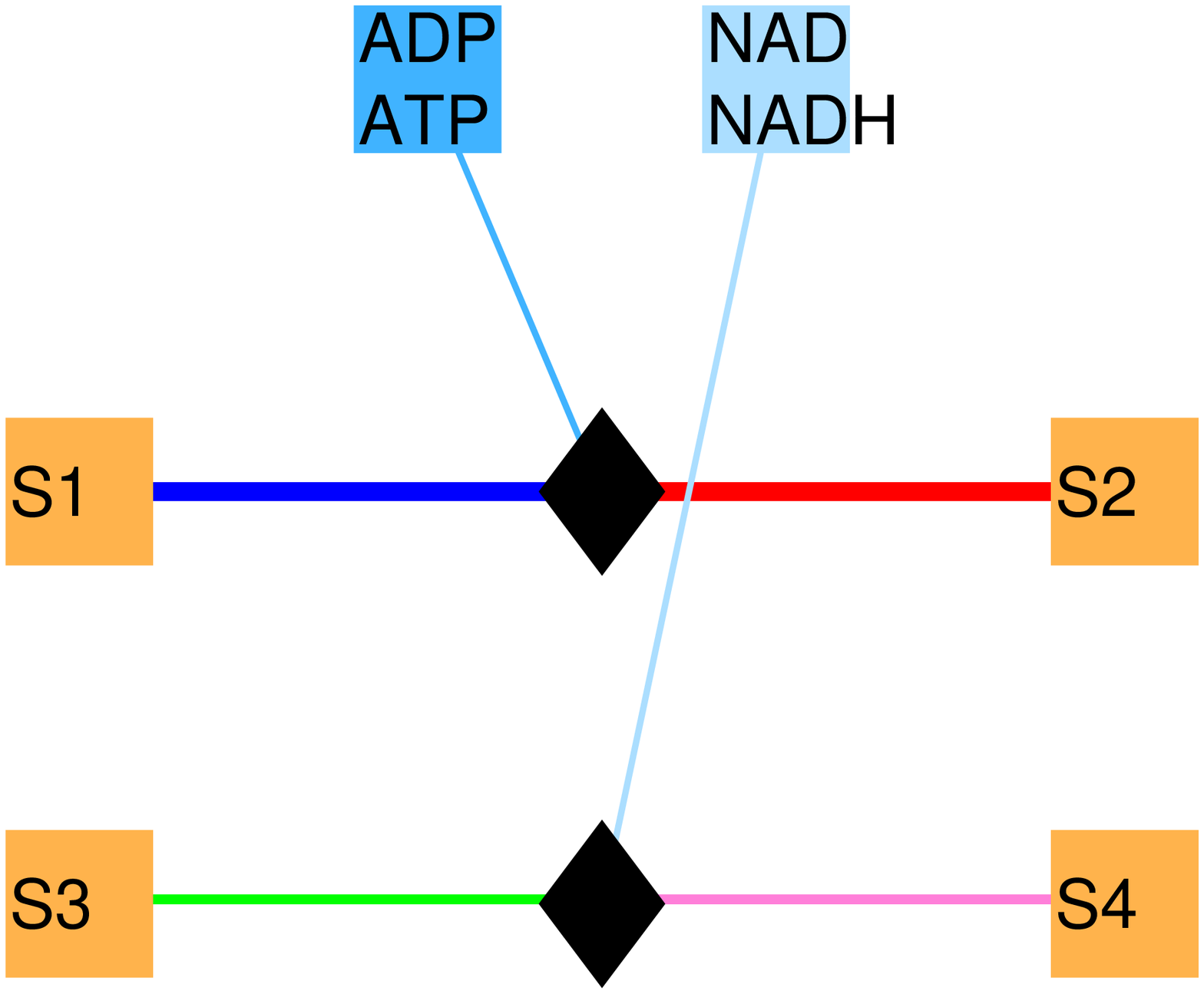}
  \end{center}
\caption{ S-graph representation of enzymatically catalyzed
reactions (\ref{ichi}) with s-matrix
(\ref{stoich_mat}). An S-graph consists of reaction nodes
(black diamond), non-carrier metabolite nodes (orange
square), and carrier metabolite nodes (light blue square).
Edges are color coded as in the s-matrix, so all the
information in the s-matrix appears schematically in the
s-graph.  This s-graph on the right is simplified by
grouping carriers which always occur in pairs (ATP/ADP,
NAD/NADH etc.). }
  \label{fig:s-graph}
\end{figure}

\begin{figure}[t]
  \begin{center}
    \includegraphics[width=\linewidth]{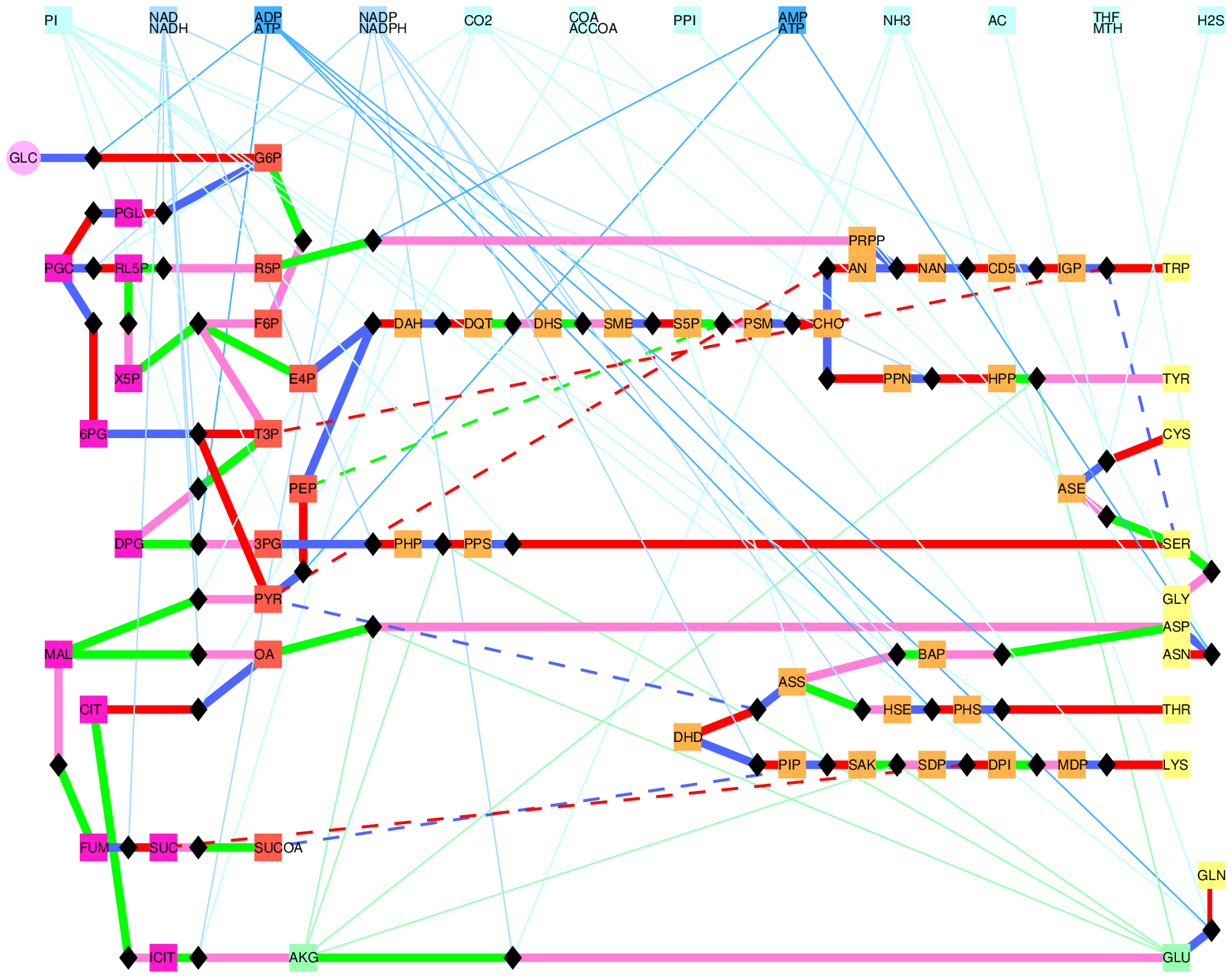}
  \end{center}
\caption{ An s-graph for the amino acid biosynthesis module
of the{\it H.Pylori} s-matrix. The conventions are same as
those in Fig.~\ref{fig:s-graph}.  This illustrates that
long biosynthetic pathways build complex building blocks
(in yellow on the right) from precursors (in orange on the
left) in a series of simple reactions (in the middle). Each
biosynthetic module has a qualitatively similar structure.}
  \label{fig:org_diag}
\end{figure}

The information conveyed in the s-matrix can be represented
in another graphical form as in Fig.~\ref{fig:s-graph} for
the simple system in (1). This is a color-coded bipartite
graph of reaction and metabolite nodes which we will call
an s-graph. The metabolite nodes can be further
differentiated into those for non-carrier and carrier
metabolites. Some carrier metabolites are always involved
in reactions as a pair and thus can be combined to simplify
the s-graph (Fig.~\ref{fig:s-graph}(right)). Models which
further reduce s-graphs to simple graphs, as is standard in
the physics literature\cite{NetBio,book}, with only either
metabolites or reactions (by elimination of the other)
destroy their biochemical meaning. All the information in
the s-matrix are conveyed to s-graphs by the same
color-coding as in the matrix with the same importance in
rows (metabolites) and columns (reactions). An example of
an s-graph is shown in Fig.~\ref{fig:org_diag} for a part
of amino acids biosynthesis module in {\it H. Pylori}.

\begin{figure*}[t]
  \begin{center}
    \includegraphics[width=0.3\linewidth,keepaspectratio,clip]{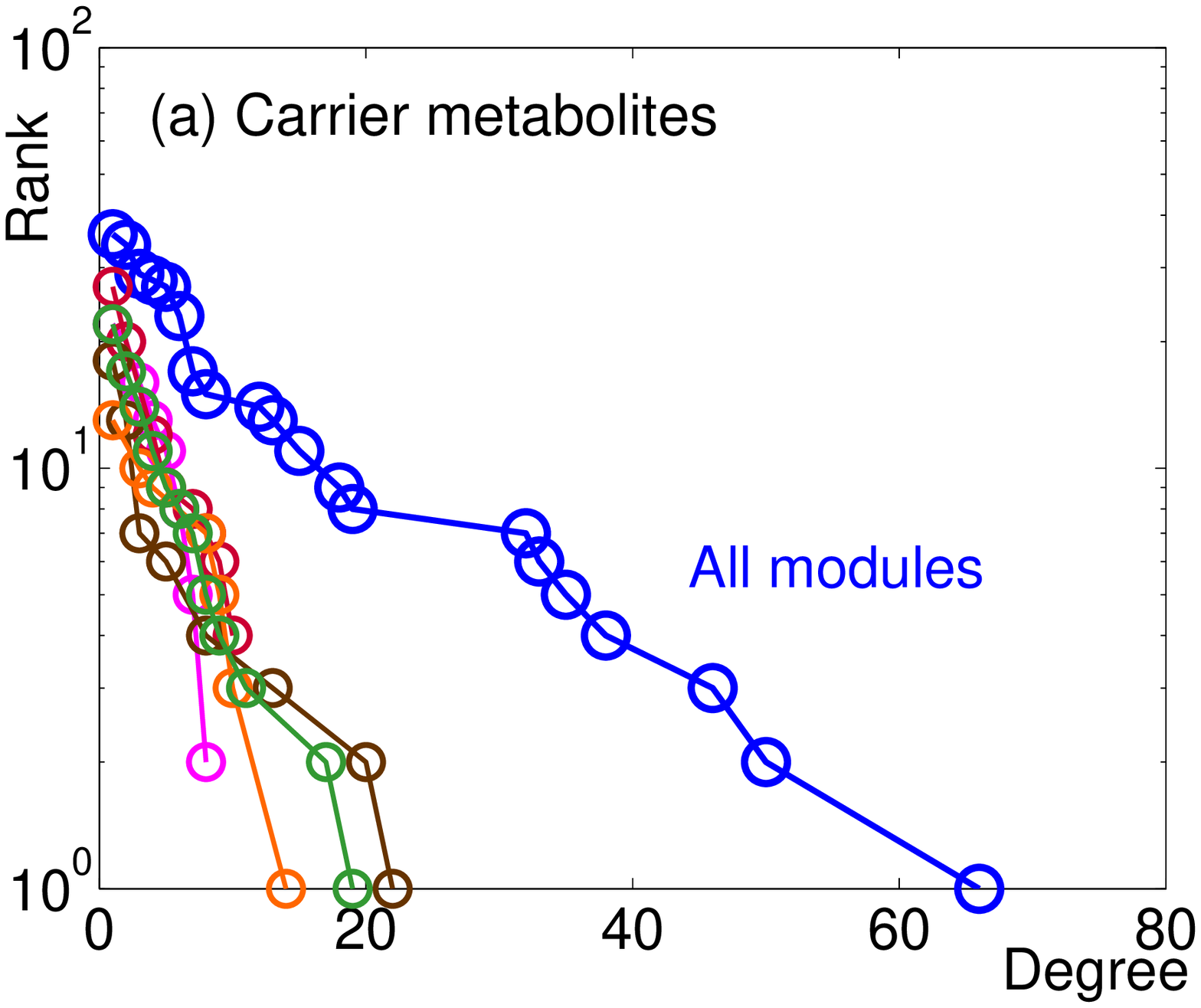}
 \includegraphics[width=0.3\linewidth,keepaspectratio,clip]{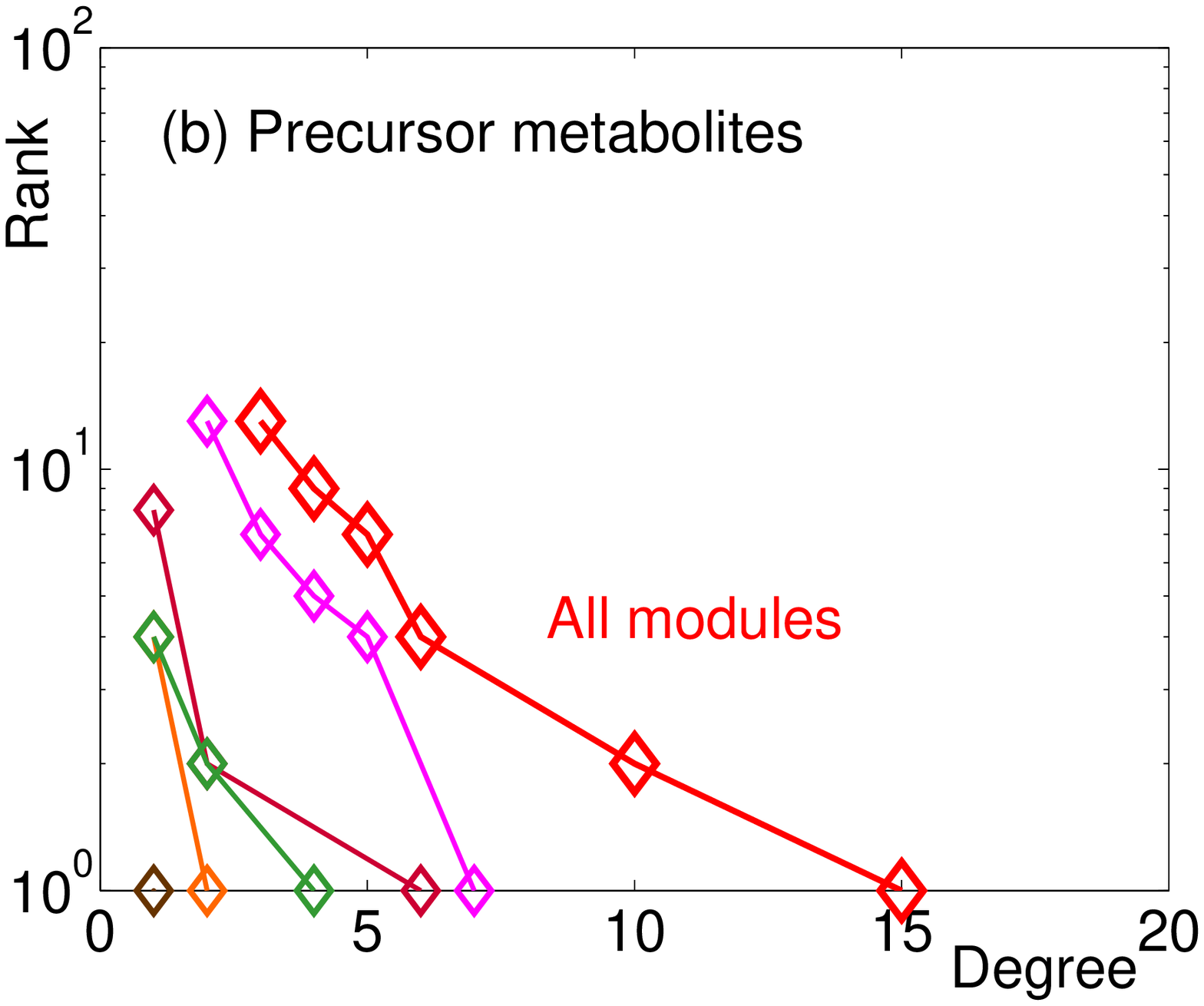}
 \includegraphics[width=0.3\linewidth,keepaspectratio,clip]{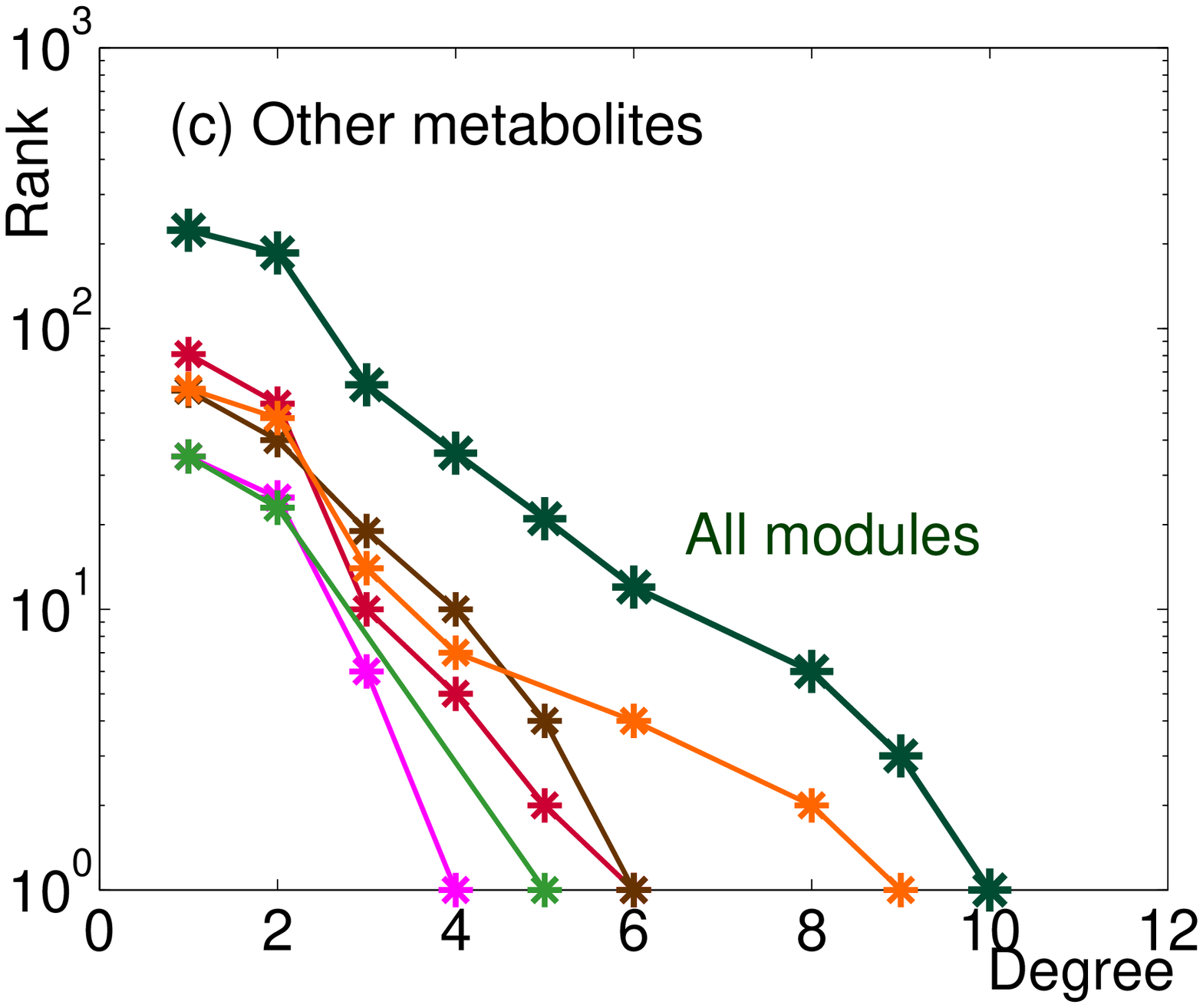}
 \includegraphics[width=0.3\linewidth,keepaspectratio,clip]{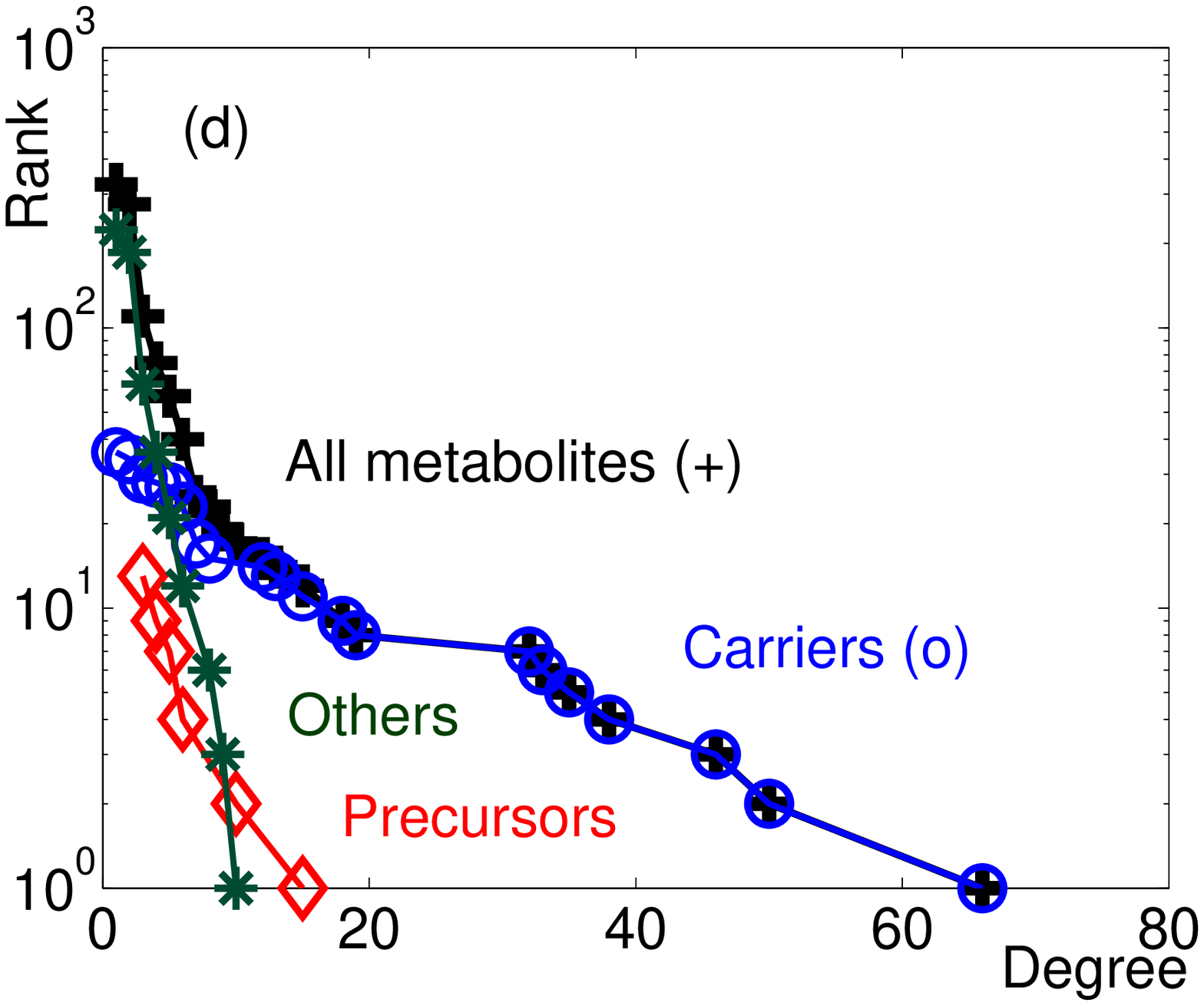}
\includegraphics[width=0.3\linewidth,keepaspectratio,clip]{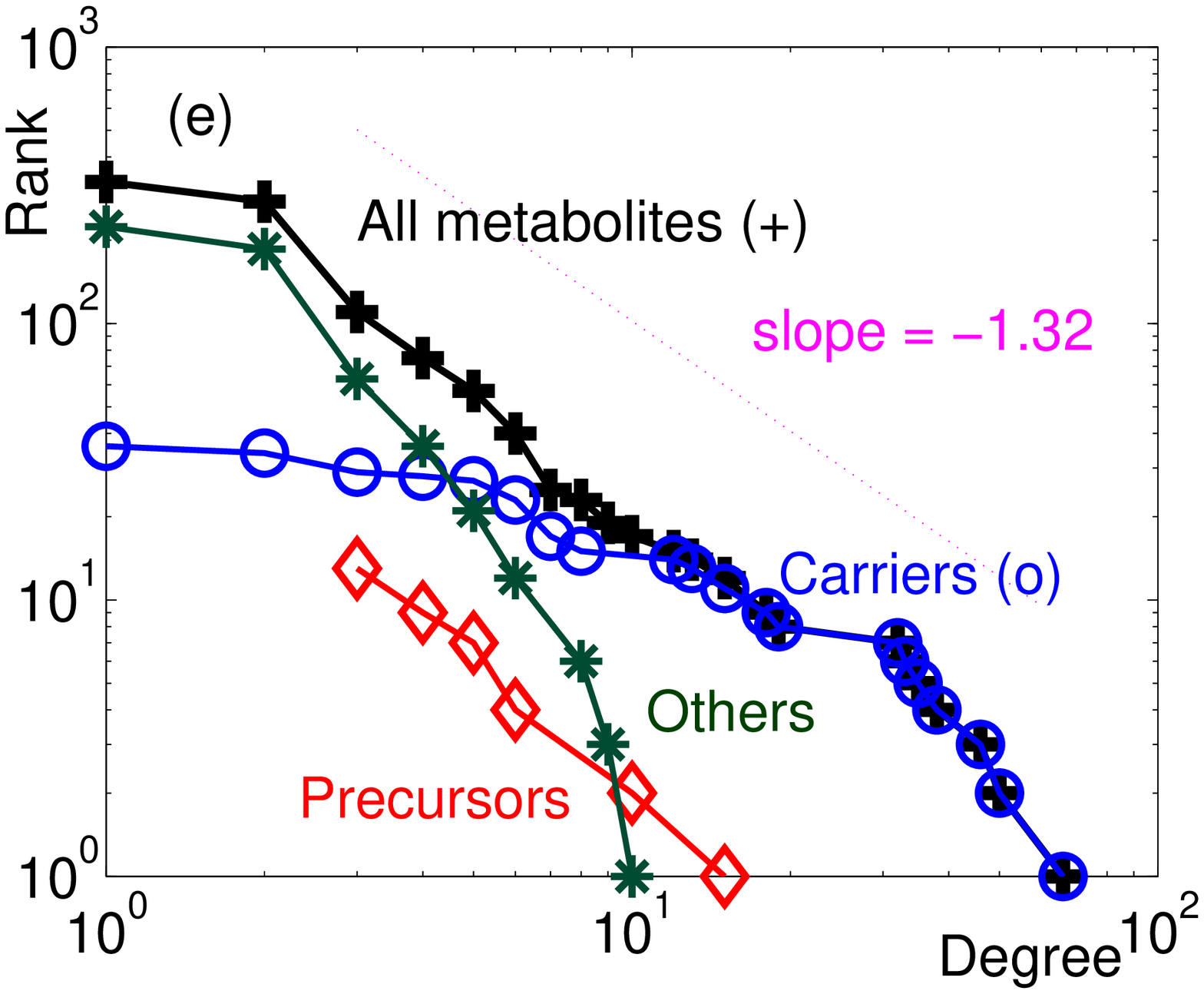}
\includegraphics[width=0.3\linewidth,keepaspectratio,clip]{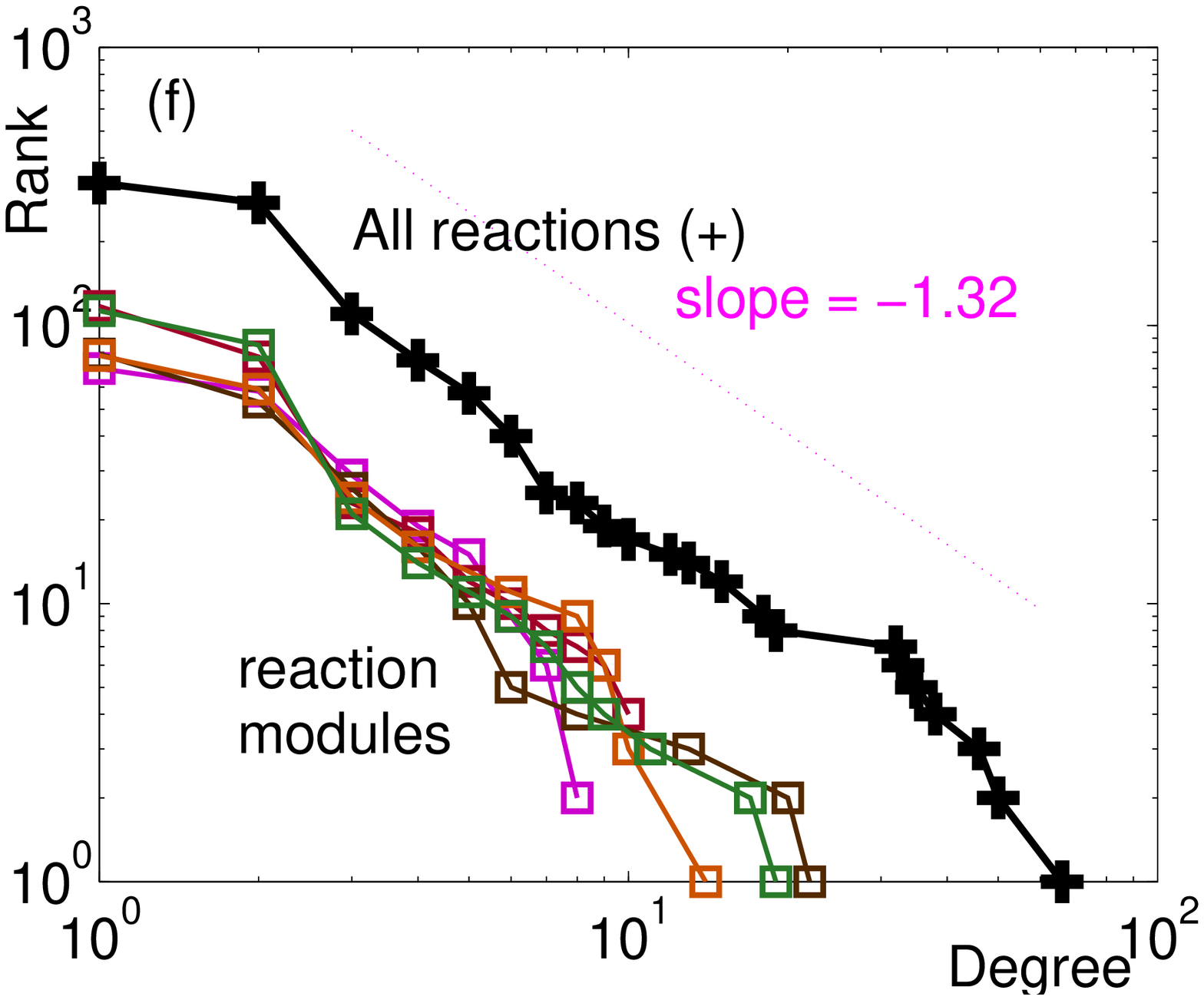}
  \end{center}
\caption{Rank (cumulative distribution) of metabolite node
degree (= number of reactions = number of links) for
metabolic networks of {\it H.Pylori}. Degrees of (a)
carrier, (b) precursor, and (c) other metabolites in the
whole network (large marker with (a) blue, (b) red, (c)
dark green) and in each reaction module (small markers
with pink, dark red, brown, orange, light green colors).
Each module shows exponential distribution. (d) Metabolite
node degrees of the whole network (black $+$) resulting
from the mixture of carrier ($\circ$), precursor
($\diamond$), and other metabolites ($\ast$), for which the
plot is the same as for (a), (b), and (c), respectively.
(e) Loglog plot of (d) indicates total degrees are
approximately power laws. (f) The total metabolites in each
reaction module with exponential distribution sums up to create the
power law distribution in the whole network.}
  \label{fig:whole_subst}
\end{figure*}

\begin{table*}[t]
\caption{Coefficients of variation of metabolite node
degree distribution. B.S: Biosynthesis. Each of carrier,
precursor, and other metabolites has low variability in
each module, and their sum results in the high variability
in total. }
   \begin{center}
    \begin{tabular}{|c||c||c|c|c|c||c|}
    \hline & Catabolism & Amino B.S. &
                Nuc. B.S. & Lipid B.S. & Vitamin B.S. & All modules\\
    \hline Others &
        0.38  & 0.49 & 0.56 & 0.67 & 0.42 & 0.61\\
    \hline Precursors &
        0.47 & 1.05 & 0 & 0.35 & 0.61 & 0.60\\
    \hline Carriers &
        0.50 & 0.81 & 1.23 & 0.64 & 0.92 & 1.13\\
    \hline\hline All metabolites &
        0.63 &  0.88 & 1.20 & 0.90 & 1.04 & 1.72\\
    \hline
          \end{tabular}
\end{center}
\label{table:var_mean}
\end{table*}

In studying degree statistics for s-graphs, degrees (number
of edges from a node) for both types of nodes, reaction and
metabolite, are important (and equivalent to degrees of
columns and rows of the s-matrix). Of particular interest
is the claim that metabolite degrees obey a power law,
which is reasonably consistent with the full network
metabolite degrees (black $+$) in
Fig.~\ref{fig:whole_subst}(d-f), which shows an approximate
power-law distribution in a log-log (e-f) rank plot, and
has clearly higher variability than an exponential as seen
in a semilog (d) plot. What is more fundamental than power
laws is high variability. For low variability processes,
Gaussians arise naturally because of the well-known central
limit theorem (CLT), and thus require no additional `special'
explanations.  Exponentials have other important invariance
properties, and are also thus quite common.  All degrees of
each module in Fig.~\ref{fig:whole_subst} are
closer to exponentials, and have low variability. Even more
important is that relaxing finite variance in the CLT
yields power laws, which are further invariant under
marginalization, mixtures, and maximization \cite{Mandel}.
Given the abundance of high variability phenomena, power
laws are an obvious null hypothesis and should properly be
viewed as `more normal than Normal'\cite{Walter}. Thus we
will focus on the mechanism responsible for low variability
in reaction and module metabolite degrees, yet high
variability in total metabolite degrees.

\begin{figure}[t]
  \begin{center}
    \includegraphics[width=0.6\linewidth,keepaspectratio,clip]{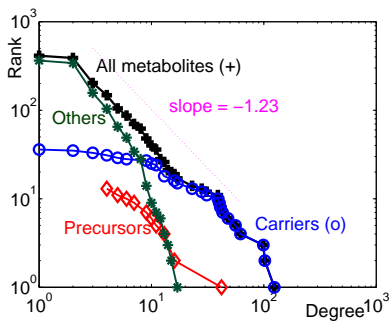}
  \end{center}
 \caption{Rank (cumulative distribution) of metabolite node
degree for {\it E.Coli} metabolism. CV$=2.05$. }
  \label{fig:Ecoli_all}
\end{figure}

Table~\ref{table:var_mean} shows the coefficient of
variation (CV$=\sigma/\mu$ where $\mu$ and $\sigma$ are
sample mean and standard deviation) for the horizontal and
vertical decomposition of the s-matrix in
Fig.~\ref{fig:whole_mat}.  The CV is a standard measure of
variability with low variability exponentials having
CV$=1$, and power laws having divergent CV for large data.
The only high variability in Table~\ref{table:var_mean}
appears for all metabolites in the full network (all
modules). It is obvious from Fig.~\ref{fig:whole_subst}
(d), which shows the decomposition of metabolites into
carrier ($\circ$), precursor ($\diamond$), and other
metabolites ($\ast$), that the high
variability in the whole network is mainly created by high
$\sigma$ from carrier metabolites mixed with low $\mu$ from
others. Figure \ref{fig:whole_subst}(a) shows the
decomposition of carrier degrees into reaction modules.
The larger marker corresponds to the degree in the whole
network, whereas the smaller ones correspond to those
in each reaction module. The sum of shared carrier
metabolites across different reaction modules pushes the total
degree of carriers much higher. In contrast, the degrees
for other metabolites ($\ast$) stays smaller
with many low degrees in total
(Fig.~\ref{fig:whole_subst}(d)). Its decomposition into
reaction modules is shown in
Fig.~\ref{fig:whole_subst}(c). As they appear almost
uniquely in each reaction module, the sum across different modules
increases the number and thus ranks, but not greatly
the degrees. The node degrees for precursor metabolites
have properties between those of carriers and others
(Fig.~\ref{fig:whole_subst}(b)). The same structure is found
in {\it E.Coli}(Fig.~\ref{fig:Ecoli_all}).

The overall high variability and thus apparent power-law is
created by a {\em mixture} of the high degree of the {\em
sum} of degrees of a few shared carriers with the many
(high-rank and) low degree of other metabolites unique to
each reaction module, with the precursors filling in between
(Fig.~\ref{fig:whole_subst}(d)).
Figure \ref{fig:whole_subst}(f) and the bottom row of
Table~\ref{table:var_mean} show another decomposition of
the all metabolites in full network ($+$) into
reaction modules, each of which has relatively low
variability. The entire network consists of widely
different scales, and thus could be called {\em
scale-rich}. 

\begin{figure}[t]
  \begin{center}
    \includegraphics[width=0.7\linewidth,keepaspectratio,clip]{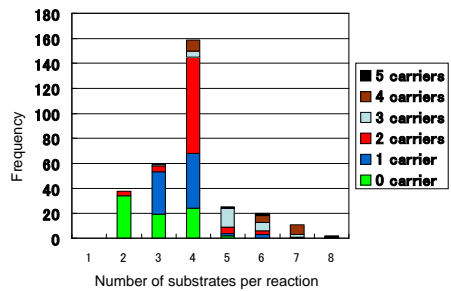}
  \end{center}
 \caption{Reaction node degree (= number of
substrates) distribution for metabolic networks of {\it
H.Pylori}. CV$=0.30$. Contributions of carrier metabolites
to degrees is also indicated. }
  \label{fig:whole_react}
\end{figure}

The reaction node degrees which are the number
of metabolites that are involved in each reaction in
Fig.~\ref{fig:whole_mat}, are shown in
Fig.~\ref{fig:whole_react}. The number of carriers involved
in a reaction is also an important statistic. The typical
reaction has four metabolites of which two are carriers,
and no reactions differ greatly from this. Overall there is
very low variability in reaction degrees, and this too can
be explained by standard biochemistry. The enzymes of core
metabolism are highly efficient and specialized for high
fluxes of small metabolites and thus necessarily have few
metabolites and involve simple reactions.  This is not
trivial, since the general purpose polymerases, chaparones,
and proteases involved elsewhere in the cell have an almost
unlimited number of distinct macromolecular substrates.

A simple model shows the essential constraints that drive
the structure of the network. In fact, the only constraint
that we will need to model from real metabolism is that
each reaction has few substrates split between shared
carriers, precursors, and others. While this organizational
structure has many additional benefits in terms of
efficiency, robustness, and evolvability\cite{Marie2}, we
will only consider how shared common carriers make high
variability at the full system level despite low
variability within metabolite modules. To emphasize the point we will
assume that each reaction has exactly one global carrier
and one other metabolite, that there is just one carrier
and it appears in every reaction, and that each other
metabolite is in just one reaction. With these assumptions,
in $r$ reactions, the $\sigma$ and therefore the CV of both
the carrier and other metabolites is exactly 0, the lowest
CV value possible. The mixture of carriers and others has
one degree $r$ carrier and $r$ degree 1 others. For large
$r$ this gives $\mu \approx 2$ and $\sigma \approx
\sqrt{r}$ so the total CV$\approx\sqrt{r}/2$. This is the
highest possible CV value that the metabolites in a
nontrivial $r$ reaction s-matrix can have.
Table \ref{table:var_sample}
shows CV for the case with $m_i$ such reactions in
module $i$ and $r=\sum m_i$.
This simple model thus shows the high variability
at the full system despite low variability within modules.

These assumptions are so extremely simplified that they
would not even allow reactions to chain together to create
pathways, but this underscores the point that the mechanism
at work here depends minimally on the properties of
metabolism per se. It simply requires a common carrier, as
is found in almost all advanced technologies, as well as
all metabolisms. Real s-matrices have broader distributions
on both metabolites and reactions and this smears out the
distributions and lowers the CV, but the qualitative
features are universal and preserved.
{\it E. Coli} has
similar modularity but more reactions than {\it H. Pylori}, and
thus a higher total CV$>2$. The strong invariance
properties of power laws means that they can be easily
caused by models based on only the most minimal constraints
of real metabolism, once they have high CV. Far from self-similar or
scale-free, these highly structured, `scale-rich,' and
self-dissimilar features of both the real data and the
simple model are the intrinsic features of metabolic
networks. The high variability is thus due to the highly
optimized and structured protocol that uses common carriers
and precursor metabolites \cite{Marie2}, and power laws are
simply the natural null statistical hypothesis for
such high variability data.  They require no
further explanation beyond this natural biological one.
\begin{table}[h]
\caption{CV for simple model.}
{\small
   \begin{center}
    \begin{tabular}{|c||c|c|c||c|}
    \hline & module$1$ & module$2$ & $\cdots$ &  All modules\\
    \hline Others &
              0  & 0   & $\cdots$ & 0\\
    \hline Carriers &
              0  & 0   & $\cdots$ & 0\\
    \hline\hline All metabolites &
           $>\sqrt{m_1}/2$ & $ >\sqrt{m_2}/2$ & $\cdots$ & $ > \sqrt{r}/2$\\
    \hline
          \end{tabular}
\end{center}
}
\label{table:var_sample}
\end{table}

In conclusion, this paper has shown that an appropriately
arranged s-matrix and its corresponding s-graph
representation enable the clear visualization of the global
bow-tie structure and reflect directly the underlying
biochemical mechanism that gives power-law metabolite node
degree distributions for the entire network.  The
decomposition of reactions and metabolites in a
biochemically meaningful way elucidates the scale-rich
structures of the network, leading naturally to power law
degree distribution for metabolite nodes. This already
shows a clear contrast between real biological networks and
models that ignore functional requirements and chemical
constraints to produce power law degree distribution
through random processes \cite{Jeong}, although this
contrast deserves further exploration and exposition.

\begin{center}
{\bf Appendix: HOT bowtie structure}
\end{center}
The robustness of the bowtie structure with a 
small knot of common currencies (carriers and precursors) is 
that it facilitates control, accommodating perturbations and 
fluctuations on many time and spatial scales. While metabolism 
allows large fluctuations in nutrients and products, relatively 
small fluctuations in ATP are lethal.  But the very architecture 
that creates this fragility also helps alleviate it, since 
ATP concentrations are tightly regulated and not easily changed. 
Another major source of fragility is that universal common currencies 
responsible for robustness are easily hijacked by parasites or 
used to amplify pathologic processes.  
Together the efficiency and adaptability of metabolism along with its 
fragilities illustrate Highly/Heterogeneous Optimized/Organized Tradeoffs/Tolerance 
(HOT)\cite{CD2}.  The metabolism bowtie architecture and associated 
protocols allow highly optimized tradeoffs between multiple requirements, 
such as reaction complexity (number of substrates in reaction), genome size, 
efficiency (energy required for each reaction), but particularly adaptability 
through tolerance of various perturbations and evolvability on longer time scales.   
Some general consequences of a HOT architecture are clear. For example, if every 
nutrient-product combination had independent pathways without shared precursors and 
carriers, the total genome would be vastly larger, and/or enzymes would be vastly more 
complex.  In either case, adaptation to fluctuating environments on any time scale would be difficult.  
Only an organization like the bowtie facilitates the kind of extreme heterogeneity that allows for 
robust regulation, manageable genome sizes, and biochemically plausible enzymes.  
\begin{table*}[ht]
    \caption{List of carrier metabolites.}
  \begin{center}
    \begin{tabular}{|l|l|}
      \hline
      Phosphate group transfer & ATP/ADP/AMP \\ \hline
      Hydrogen transfer & NADH/NAD, NADPH/NADP, FADH/FAD, 
				OTHIO/RTHIO, MK/MKH$_2$ \\ \hline
      Amino group transfer & AKG/GLU \\ \hline
      Acetyl group transfer &
      ACCOA/COA \\ \hline
      One carbon unit transfer & THF/METTHF/FTHF/MTHF/METHF  \\\hline 
      Others & CO$_2$,NH$_3$,O$_2$,H$_2$O$_2$,
      H$_2$CO$_3$
      H$_2$S,
      H$_2$SO$_3$,
      NO$_2$ \\ 
      & Sulfate,
      Acetate,
      H$^+$,
      Phosphate,
      Pyrophosphate, ACP \\ \hline
    \end{tabular}
\end{center}
\label{carriers}
\end{table*}


{\small

}


\end{document}